\documentclass[twocolumn,letterpaper]{IEEEAerospaceCLS}  
\usepackage[pdftex]{graphicx}

\DeclareGraphicsExtensions{.png,.pdf,.jpg,.eps}

\usepackage{amsmath}
\usepackage{amsfonts,amssymb}

\def\bF{{\mathbb{F}}}

\renewcommand{\P}[1]{{\mathbb{P}}\left(#1\right)}

\newcommand{\ignore}[1]{}  

\begin{document}

\title{Information-theoretic Physical Layer Security for Satellite Channels}
\author{
\'{A}ngeles Vazquez-Castro\\
Department of Telecommunications\\ and Systems Engineering,\\
Autonomous University of Barcelona\\
 \and
Masahito Hayashi\\
Graduate School of Mathematics, Nagoya University and\\
Centre for Quantum Technologies, \\National University of Singapore\\

\maketitle

\thispagestyle{plain}
\pagestyle{plain}

\begin{abstract}
Shannon introduced the classic model of a cryptosystem in 1949, where Eve has access to an identical copy of the cyphertext that Alice sends to Bob. Shannon defined perfect secrecy to be the case when the mutual information between the plaintext and the cyphertext is zero. Perfect secrecy is motivated by error-free transmission and requires that Bob and Alice share a secret key. Wyner in 1975 and later I.~Csisz\'ar and J.~K\"orner in 1978 modified the Shannon model assuming that the channels are noisy and proved that secrecy can be achieved without sharing a secret key. This model is called wiretap channel model and secrecy capacity is known when Eve's channel is noisier than Bob's channel. 

In this paper we review the concept of wiretap coding from the satellite channel viewpoint. We also review subsequently introduced stronger secrecy levels which can be numerically quantified and are keyless unconditionally secure under certain assumptions. We introduce the general construction of wiretap coding and analyse its applicability for a typical satellite channel. From our analysis we discuss the potential of keyless information theoretic physical layer security for satellite channels based on wiretap coding. We also identify system design implications for enabling simultaneous operation with additional information theoretic security protocols.  
\end{abstract}

\tableofcontents

\section{Introduction}
\label{sec:fund-mech-inform}
Traditionally, communication transmission for civilian space missions have not been employing security mechanisms. Moreover, safety is considered prevalent over security in civilian space missions, which makes the trade-off between safety and security a specific design feature of space missions. Space agencies are fostering civilian cryptography research to develop schemes for security enforcement which can be recommended by relevant standardization bodies such as the Consultative Committee for Space Data Systems (CCSDS) \cite{Ignacio2010}. The current Space Data Link Security (SDLS) protocol can provide security services to the existing CCSDS family of Space Data Link (SDL) protocols: Telecommand (TC), Telemetry (TM) and Advanced Orbiting Systems (AOS). CCSDS does not mandate at which layer the encryption algorithm is used. Security can also be implemented at the upper layers or simultaneously at multiple layers.

On the other hand, civilian commercial missions, have primarily focused on improving system capacity, Quality of Service (QoS) and of Experience (QoE). Successful standards such as the Digital Video Broadcasting Standard DVB-S2X enables multiple interoperable DVB-S2 ecosystems. This is because the physical layer is highly modular with powerful modulation and coding schemes. Notably, confidentiality, integrity or data origin authentication must be taken care of at the upper layers through e.g. standard internet security protocols. However, it is well known that such protocols through e.g. tunnelling may lead to significant transmission overhead in clear detriment of QoS. Furthermore, ecosystems implementing other known protocols are prone to their known corresponding attacks.

The engineering approach to the design of physical layer security mechanisms of space links mostly relies on low probability of intercept (LPI) and low probability of detection (LPD) waveforms. LPI and LPD waveforms along with signal processing and channel coding techniques are highly effective against jamming attacks and interception. Their anti-jamming performance can be quantified in terms of the power level and signal at which the protection is broken. As for interception, security performance  can be quantified in terms of breaking the pseudo-random sequence spreading the signal. However, novel attacks and computing power evolution could end up breaking their computationally strong security.

In this paper we introduce the information-theoretic approach to the design of keyless physical layer security for space links. This method exploits physical characteristics of the communication channel and can numerically quantify secrecy according to information-theoretic principles. This allows the design of cryptographic coding schemes that can lead to practical schemes adapted to channel, system and mission specific constraints and design goals. The first study on the applicability of information-theoretic physical layer security to space links was \cite{Lei2011} where it is proposed joint power and multi beam antenna weights optimisation to meet individual secrecy rate constraints. This work as well as subsequent works \cite{Kalantari2015} apply signal processing techniques rather than wiretap coding and their focus is weak secrecy level. In this paper we take a step forward focusing on evaluating the potential of one-way keyless information theoretic wiretap coding. Th
 e aim is
  to provide both reliability and cryptographic semantic security.

The main contributions of this positioning paper are three. First, we clarify in Section II the information-theoretic framework of physical layer security based on wiretap coding. We describe the original wiretap channel and its generalisation to include cryptographic semantic level of security. Second, we introduce the general construction of wiretap codes to realise physical layer security as well as a randomised hash function construction, for which we show theoretical performance. 
Finally, we illustrate the realistic potential of wiretap coding for satellite channels through analysis and simulations that provide quantitative results of secrecy performance over existing satellite systems. 
We also identify system design implications for possible simultaneous operation with additional information theoretic security mechanisms.  

\section{The wiretap channel and security criteria}
\label{sec:security-criteria}

\subsection{The wiretap channel}

Information-theoretic security can be traced back to Shannon, who introduced the classic model of a cryptosystem in 1949, where Eve has access to an identical copy of the cyphertext that Alice sends to Bob. Shannon defined perfect secrecy to be the case when the plaintext and the cyphertext are statistically independent. Perfect secrecy is motivated by error-free transmission and requires that Bob and Alice share a secret key.

A. Wyner in 1975~\cite{Wyner1975} relaxes the stringent condition of perfect statistical independency of Shannon's cryptosystem and introduced the wiretap channel model. In this model, transmission is not error-free. The wiretap channel model is composed of two channels. One is the channel from the legitimate transmitter (Alice) to the legitimate receiver (Bob), which is referred to as the ``main'' channel, and is considered to be a memoryless channel characterized by input alphabet ${\cal X}$, output alphabet ${\cal Y}$, a transition probability $W_{Y |X}$. The other is the channel from Alice to a passive adversary (Eve), which is referred to as the ``eavesdropper's channel'', and consists of another memoryless channel characterized by input alphabet ${\cal X}$,  output alphabet ${\cal Z}$, and transition probability $W_{Z|X}$.
This model supposes that the statistics of both channels are known to all parties, and that authentication is already done\footnote{A small secret key is needed to authenticate the communication. It is known that $\log n$ bits of secret are sufficient to authenticate $n$ bits of data~\cite{Wegman1981}}. Also, it is assumed that Eve knows the coding scheme used by Alice and Bob.

The channel can be probabilistic or combinatorial. The probabilistic channel (known as Type I) is given in terms of the transition probabilities $W_{Y |X}$ and $W_{Z|X}$. The combinatorial channel (known as Type II) is given as an adversarial channel model, where the intruder is allowed to observe $\mu\leq N$ components of $X^{N}$. In either case, wiretap code design goal is the simultaneous provision of reliability and security. 

In~\cite{Wyner1975} Wyner considered the special case where both the main and the eavesdropper channel are discrete memoryless channels (DMCs). Further, under the assumption that the eavesdropper channel is degraded with respect to the main channel, he defined and obtained a unique parameter characterising the wiretap system, the secrecy capacity. This parameter means that for $\epsilon > 0$, there exist coding schemes that can provide secure and reliable rate (secrecy rate) $R_{s}>C_{s}-\epsilon$. Wyner proved that Alice can send information to Bob in perfect secrecy over a noisy channel without sharing a secret key with Bob initially. 

Wyner's original wiretap channel model was generalised by I.~Csisz\'ar and J.~K\"orner, in 1978~\cite{Csiszar1978} who considered the information-theoretic discrete memoryless broadcast channel. The noisy broadcast channel is characterised by the conditional distribution $W_{YZ |X}$ so that Wyner's model corresponds to the special case where $V\rightarrow X\rightarrow YZ$ is a Markov chain and $P_{YZ |X}$  factors as $W_{Y |X}W_{Z |X}$. The secrecy capacity $C_{s}(W_{YZ |X})$ is the maximal rate at which Alice can reliably send information to Bob such that the rate at which Eve obtains this information is arbitrarily small. The secrecy capacity is positive whenever the legitimate channel has an advantage in terms of the broadcast channel's conditional distribution  $W_{YZ |X}$.  

The framework was later generalised to meet cryptographic symmetry-breaking methods by~\cite{Maurer1993} and~\cite{Bellare2012}. The cryptographic generalisation identifies the logical equivalence between some cryptographic and information theoretic security metrics while introduces additional ones not used in Wyner's model. This cryptographic treatment of the wiretap channel induces fundamental differences in terms of design goals of the resulting cryptographic wiretap system. First, the provision of security can be decoupled from the design of codes for reliability. Second, the system departs from the one-way Shannon transmission system model and enables implementing different security protocols. 

The cryptographic view of the wiretap channel can also be framed within spectrum information-theoretic methods, as established by Hayashi \cite{hay-wire}. This framework provides a methodology for the design of wiretap codes based on the notion of channel resolvability.

\subsection{Security metrics and wiretap coding design goals} 

Different security metrics correspond to different wiretap code design goals. It would seem natural to relate the secrecy of the message $M\in {\cal M}_n$ to the probability of decoding error from the observation $Z^n$~\cite{Klinc2011,Baldi2012}. However, error probability depends on a distance metric between $M$ and $Z^n$ while the desired security metric is related to the information content in $Z^n$. We provide below security metrics in increasing strength order and the corresponding design criteria of wiretap codes.


\textbf{\emph{Weak secrecy.}} 
This criterion was introduced by Wyner \cite{Wyner1975} and can be given by the normalized mutual information rate
  \begin{align}
    \label{eq:weak_secrecy}
    \text{S}_\text{weak}({M};Z^n)=\frac{1}{n}I({M};Z^n) = \frac{1}{n}{\mathbb{D}}(p_{{M}Z^n}\| p_{M}p_{Z^n}),
  \end{align}
where $\mathbb{D}$ is the Kullback-Leibler divergence. 

A wiretap code under this security metric should simultaneously provide reliability and security. 
\begin{itemize}
\item \textbf{Reliability condition}: $\underset{n\rightarrow\infty}{\lim} 
\P{\hat{M}\neq M}=0$. 
\item \textbf{Security condition}: $\underset{n\rightarrow\infty}{\lim}\text{S}_\text{weak}({M};Z^n)=0$. 
\end{itemize}

Note that in this case wiretap code design follows classic coding theory. For the probabilistic Type I wiretap channel, the secrecy rate at which Alice and Bob can reliably and securely communicate will tend, with increasing block length, to the maximum possible value given by the secrecy capacity. 
The results with weak security were extended to the Gaussian channels in \cite{Leung-Yan-Cheong1978}. 

For the combinatorial Type II wiretap channel, codes are designed to provide perfect secrecy in terms of known algebraic properties over hamming spaces. 
There are known secrecy capacity-achieving constructions under this security metric based on low density parity (LDPC) codes \cite{TDCMJ} and polar codes \cite{MV}.

This security metric was strengthened as follows.


\textbf{\emph{Strong secrecy.}}
The weak security metric criterion was strengthened from a security point of view by subsequent improvement of Wyner's model ~\cite{Maurer1994} and consists in characterizing the total information leakage
  \begin{align}
    \label{eq:strong_secrecy}
    \text{S}_\text{strong}({M};Z^n)
=I({M};Z^n) = 
{\mathbb{D}}(p_{{M}Z^n} \Vert p_{M}p_{Z^n}).
  \end{align}
A wiretap code under this security metric should simultaneously provide reliability and security and was given in \cite{csiszar96,hay-wire}.  
\begin{itemize}
\item \textbf{Reliability condition}: $\underset{n\rightarrow\infty}{\lim}   \P{\hat{M}\neq M}=0$. 
\item \textbf{Security condition}: $\underset{n\rightarrow\infty}{\lim}{    \text{S}_\text{strong}({M};Z^n)
}=0$. 
\end{itemize}

Note that wiretap code design for both weak and strong secrecy is according to classic coding theory. 
Specifically, for fixed rate, a wiretap coding scheme (encoding and decoding algorithms) is a family of increasing block length codes that provide either weak or strong security conditions.
For the Gaussian channels, a wiretap code under this security metric was given in \cite[Appendix D-C]{Hayashi2012}
When we use LDPC codes,
the security under this security metric was shown only for 
the binary erasure wiretap channel \cite{STBM},
Also, there are known secrecy capacity-achieving constructions under this security metric
based on polar codes \cite{CBK}.

\textbf{\emph{Semantic secrecy.}} This information theoretic secrecy criterion was formalized by Bellare, Tessaro, and Vardy~\cite{Bellare2012,Bellare2012arXiv} by adapting the notion of semantic secrecy used in computational cryptography~\cite{Goldwasser1984}. In this setting, classic coding theoretic design and terminology is replaced by cryptographic methods and terminology. In this setting, a wiretap coding scheme is generalised to an encryption scheme, a family of encryption functions (encryption and decryption algorithms).

Semantic secrecy generalises the previous secrecies, which are included under particular conditions. This is achieved by introducing a fundamental difference on how to measure secrecy, which is now measured in terms of indistinguishability. In addition, the design conditions of semantic secrecy systems are independent of the receiver channel. Further, the semantic security metric is measured in terms of a quantitative real number, the advantage $\text{Adv}({M};Z^n)$, defined as
  \begin{align}
&\text{Adv}({M};Z^n) \nonumber \\
=&\max_{f,p_{M}}\bigg(\sum_{z^n}p_{Z^n}(z^n)\max_{f_i\in\text{supp}(f)}\P{f({M})=f_i|z^n}\nonumber \\
&\quad -\max_{f_i\in\text{supp}(f)}\P{f({M})=f_i}\bigg).
    \label{eq:advantage_secrecy}
  \end{align}
where $f$ is an additional channel with domain the message set that represents partial information about the message. Accordingly, the following distinction is in place \cite{Bellare2012}:

\begin{itemize}
\item \textbf{Qualitative secrecy condition}: This condition establishes only asymptotically whether or not a encryption scheme is secure. It is defined as $\underset{n\rightarrow\infty}{\lim}{\text{Adv}({M};Z^n)}=0$. However, this does not mandate any particular rate at which the advantage should vanish.
\item \textbf{Quantitative secrecy condition}: This condition establishes a numerical secrecy metric as measured by $\text{Adv}({M};Z^n)$.
\end{itemize}

Essentially, $\text{Adv}({M};Z^n)<\epsilon$ means that the knowledge of the observation $Z^n$ increases the probability of guessing any function $f$ of the message ${M}$ by no more than $\epsilon$, irrespective of its distribution $p_{M}$. For example, an encryption function is said to achieve $s$-bit semantic security if $\text{Adv}({M};Z^n)<2^{-s}$.


Under such distinction, it is possible to establish conditions for the logical equivalence between semantic and strong secrecy. In particular, semantic secrecy can be guaranteed by upper bounding the quantity 
$\max_{p_{M}}\text{S}_{\text{strong}}({M};Z^n)$ with a proper coefficient.
Since the security criterion 
$\max_{p_{M}}\text{S}_{\text{strong}}({M};Z^n)$
expresses the strong security independent of the source distribution $P_{M}$,
it is called the source universal security \cite[Section XIII]{Hayashi2012}. 

\subsection{Security analysis}
 
The above criteria and conditions serve as design goals of wiretap codes that can provide required levels of reliability and security to a communication system or service.

The security analysis of specific designs of wiretap codes needs to comply with the wiretap coding assumptions of operation,
which are 
\begin{description}
\item [{(1)}] communication is one-way, 
\item [{(2)}] the adversary is computationally unbounded, 
\item [{(3)}] the security scheme is keyless, 
\item [{(4)}] the channel of the adversary is noisier than the legitimate
channel. 
\end{description}
Note that condition  $\left(3\right)$ assumes that sender and receiver do not share \textit{a priori} any information not known to the adversary. Also note that $\left(4\right)$ depends on the wiretap system model.

\vspace{5 mm}
In the remainder of the paper, 
an $\left(n,k,\epsilon,\delta\right)$ code refers to a code for the wiretap channel that transmits one of $2^k$ distinct messages over $n$ uses of the channel, 
ensures an average probability of error less than $\epsilon$, and guarantees a secrecy leakage measured in terms of $\text{S}_{\text{strong}}$ or $\text{Adv}$ less than $\delta$. 
The quantity $k/n$ is the \emph{rate} of the code.

Most information-theoretic studies of the wiretap channel focus on characterizing the \emph{secrecy capacity} $C_s$, 
which is the maximum of the transmission rate $k/n$ with arbitrarily low probability of error and secrecy leakage in the limit $n \to \infty$. 

Formally, $C_s$ is the supremum of all rates $R$, such that there exists sequences of $(n,k_n,\epsilon_n,\delta_n)$ codes with
\begin{align*}
  \lim_{n\rightarrow\infty}\frac{k_n}{n}\geq R,\quad \lim_{n\rightarrow\infty}\epsilon_n=\lim_{n\rightarrow\infty}\delta_n=0.
\end{align*}
The secrecy capacity of a discrete memoryless wiretap channel 
is completely characterized for the weak secrecy, strong secrecy, and semantic secrecy metrics, as~\cite{Csiszar1978}
\begin{align}
  \label{eq:secrecy_capacity}
  C_s = \max_{p_{VX}}\left(I(V;Y)-I(V;Z)\right)_+,
\end{align}
where
$I(V;Y)$ and $I(V;Z)$ 
express the mutual informations under the distributions
$p_{VXY}:=W_{Y|X}p_{XV}$
and $p_{VXZ}:=W_{Z|X}p_{XV}$,
and 
$(a)_+$ is $a$ for a positive number $a$ and is $0$ for a negative number $a$.

In particular, $C_s$ is positive when the eavesdropper's channel is not less noisy than the main channel~\cite{Csiszar1978}, i.e., unless 
the inequality $I(V;Z)\geq I(V;Y)$ holds for any joint distribution $p_{VX}$.

Such characterization of the $C_s$ and corresponding achievability corresponds to qualitative secrecy conditions. The analysis of quantitative secrecy conditions, corresponds to the case where the wiretap codeword length, $n$, is just large enough to achieve
a given target reliability. For this reason it is denoted as analysis in the finite-length regime. The intuition for this regime is that the codeword length is large enough to achieve a target reliability and/or secrecy performance. In this case the communication system will not be able to filter out the stochastic variability introduced by the actual channel. The secrecy analysis in the finite-length regime boils down to characterize the number of semantic bits that can be transmitted. This number will provide intuition on the relations and tradeoffs  between system design performance targets and achievable secrecy. 

\section{General wiretap code construction}
Now, we discuss how to make a wire-tap channel code to realize the security.
First, we prepare an auxiliary random variable $L_n$ subject to the uniform distribution on another set ${\cal L}_n$.
Then, we prepare an error correcting code.
That is, we prepare an encoder as a map 
$\phi_{e,n}$ from 
the set 
${\cal M}_n \times {\cal L}_n$ to the input alphabet ${\cal X}^n$,
and a decoder as a map $\phi_{d,n}$ from 
the input alphabet ${\cal X}^n$ to the set ${\cal M}_n \times {\cal L}_n$.

We also prepare 
another map $f_n $ from
${\cal M}_n \times {\cal L}_n$ to ${\cal M}_n$
satisfying the condition
\begin{align}
|f_n^{-1}(m)|=| {\cal L}_n|
\end{align}
for $m \in {\cal M}_n$. This map is often called a hash function.
The encoder of our wire-tap code is given as follows.
When Alice intends to send a message $m$, 
using the auxiliary random variable $L_n$,
she generates a random variable $L_n'$ on $f_n^{-1}(m)$
because the cardinality of $f_n^{-1}(m)$ is the same as that of ${\cal L}_n$.
Then, she sends the alphabet $ \phi_{e,n} (L_n')$.
The decoder is given as the map $f_n \circ \phi_{d,n}$.

When ${\cal X}$ has a modular structure and 
the channels $W_{Y|X}$ and $W_{Z|X}$ satisfy covariant properties,
the channel is called symmetric.
Now, we consider the special symmetric case when 
${\cal X}=\{0,1\}=\bF_2$.

Now, we employ an algebraic error correcting code for the case with $ {\cal X}=\bF_2 $.
That is,
the map $\phi_{e,n}$ is given as a homomorphism from 
the set 
${\cal M}_n \times {\cal L}_n$ to the input alphabet ${\cal X}^n=\bF_2^n$,
where ${\cal M}_n$ and $ {\cal L}_n$ are given as $\bF_2^{k_{n}}$ and $\bF_2^{k_{n}'}$,
and 
the map $ f_n$ from
${\cal M}_n \times {\cal L}_n$ to ${\cal M}_n$ is given as an isomorphism.

Such an algebraic code is given as a pair of two sublinear spaces $ C_2\subset C_1 \subset \bF_2^n
$, where $C_1$ and $C_2$ are isomorphic to $\bF_2^{k_{n}+k_{n}'}$ and $\bF_2^{k_{n}'}$, respectively.
So, the map $\phi_{e,n}$ is given as the isomorphism from $\bF_2^{k_{n}+k_{n}'}$ to $C_1$.
Using the isomorphic from the quotient space $C_1/C_2 $ to $\bF_2^{k_{n}}$,
we define the map $ f_n$ from ${\cal M}_n \times {\cal L}_n$ to ${\cal M}_n$.
This type of encoder is called coset encoding.

\section{Randomized hash function}
\label{sec:chann-resolv-with}

In this section, we fix an algebraic error correcting code $(\phi_{e,n},\phi_{d,n})$.
Here, we choose integers $k_n$ and $k_n'$ such that
$k_n+k_n' $ is the message length of the given algebraic error correcting code $(\phi_{e,n},\phi_{d,n})$ and 
$k_n$ is the message length of our secrecy transmission.
So, $k_n'$ can be regarded as the sacrifice bit-length in our protocol.
Then, we consider a randomized hash function $F:\bF_2^{k_n+k_n'}\rightarrow\bF_2^{k_n}$
satisfying the following property:
\begin{align}
  \label{eq:condition_masahito}
\forall m\neq \forall m' \in\bF_2^{k_n+k_n'}
\quad
\P{F(m)=F(m')}\leq\frac{1}{2^{k_n}}.
\end{align}
This condition is called univeral2\cite{Wegman1981,Carter}, and will be supposed in the remaining part.
A modified form of the Toeplitz matrices is also shown to be universal$_2$, 
which is given by a concatenation $(T(S), I)$ of the 
$k_n' \times k_n$ Toeplitz matrix $T(S)$ and the $k_n \times k_n$ identity matrix $I$ \cite{Hayashi2011},
where $S$ is the random seed to decide the Toeplitz matrix and belongs to $\bF_2^{k_n+k_n'-1}$.
The (modified) Toeplitz matrices are particularly useful in practice, because there exists an efficient multiplication algorithm using the fast Fourier transform algorithm with complexity $O(n\log n)$. 

When the random seed $S$ is fixed, 
the encoder for our wire-tap code is given as follows.
By using the auxiliary random variable $L_n \in \bF_2^{k_n'}$,
the encoder is given as $\phi_{e,n}
\Big( 
\Big(
\begin{array}{cc} 
I & - T(S) \\
0 & I
\end{array}
\Big)
\Big(
\begin{array}{c} 
M \\
L_n
\end{array}
\Big)
\Big)$
because 
$
(I,T(S))
\Big(
\begin{array}{cc} 
I & - T(S) \\
0 & I
\end{array}
\Big)
= (I, 0)$.
(Toeplitz matrix $T(S)$ can be constructed as a part of circulant matrix. 
For example, the reference \cite[Appedix C]{Hayashi-Tsurumaru}
gives a method to give a circulant matrix.).
More efficient construction for univeral2 hash function is discussed in \cite{Hayashi-Tsurumaru}.

So, the decoder is given as $Y^n \mapsto (I,T(S)) \phi_{d,n}(Y^n)$.
When $\phi_{d,n}$ can be efficiently performed like a LDPC code or a Polar code,
That is, as long as the algebraic error correcting code $(\phi_{e,n},\phi_{d,n})$
can be efficiently performed our code can be efficiently performed.

Using the channel to eavesdropper described by a transition matrix $W_{Z|X}$,
we define the function
\begin{align}
&\psi(s|W_{Z|X},q_X)\nonumber \\
:=& \log \sum_{x,z} q_X(x) W_{Z|X}(z|x)^{1+s}
(\sum_x q_X (x) W_{Z|X}(z|x))^{-s} .
\end{align}
This function satisfies 
\begin{align}
\frac{\psi(s|W_{Z|X},q_X)}{s}\to I(X;Z)
\end{align}
as $s \to 0$.

Firstly, for simplicity, we consider the case when the main channel $W_{Y|X}$ is noiseless
and the channel to the eaves dropper is given as the $n$-fold extension of
$W_{Z|X}$.
In this case, the error correcting encoder $\phi_{e,n} $ is the identity map and $k_n+k_n'=n$.
So, as a special case of \cite[(12)]{Hayashi2011},
the above given code satisfies 
\begin{align}
{\mathbb{E}}_F e^{s \text{S}_{\text{strong}}(M;Z^n)} 
\le 1+ 2^{-s k_n'}e^{n\psi(s|W_{Z|X},q_X)}
\end{align}
for $s \in [0,1]$.
Jensen inequality yields that
\begin{align}
{\mathbb{E}}_F  \text{S}_{\text{strong}}(M_n;Z^n)
\le &\frac{1}{s} \log (1+ 2^{-s k_n'}e^{n \psi(s|W_{Z|X},q_X)}) \nonumber \\
\le &\frac{1}{s} 2^{-s k_n'}e^{n \psi(s|W_{Z|X},q_X)}
\end{align}
for $s \in [0,1]$.
So, when $\frac{\psi(s|W_{Z|X},q_X)}{s}$ is smaller than $\frac{k_n}{n} \log 2$,
the average of the leaked information 
${\mathbb{E}}_F  \text{S}_{\text{strong}}(M_n;Z^n)$ goes to zero exponentially.

Now, we proceed to the general case, i.e., the case when
the main channel $W_{Y|X}$ is noisy.
So, 
the error correcting encoder $\phi_{e,n} $ is not the identity map, and
$k_n+k_n'$ is smaller than $n$.
To discuss the general case, we introduce other functions
\begin{align}
E_0(s|W_{Z|X},q_X):=& \log 
\sum_{z}
(\sum_x q_X(x) W_{Z|X}(z|x)^{\frac{1}{1-s}} )^{1-s} \\
E_{0,\max}(s|W_{Z|X}):=& \max_{q_X}E_0(s|W_{Z|X},q_X).
\end{align}
This function satisfies 
\begin{align}
\frac{E_0(s|W_{Z|X},q_X)}{s}\to I(X;Z)
\end{align}
as $s \to 0$.
When $W_{Z|X}$ is symmetric,
$E_{0,\max}(s|W_{Z|X})$ equals $E_0(s|W_{Z|X},q_X)$ with uniform distribution $q_X$ for $X$.

The above given code
\cite[Lemma 1, Theorem 4, (16)]{Hayashi2011} satisfies 
\begin{align}
{\mathbb{E}}_F  \text{S}_{\text{strong}}(M;Z^n)
\le \frac{1}{s} 2^{-s k_n'}e^{n E_{0,\max}(s|W_{Z|X})}. \label{H12-13}
\end{align}
for $s \in [0,1]$.
So, when $\frac{E_{0,\max}(s|W_{Z|X})}{s}$ is smaller than $(\frac{k_n'}{n})\log 2$,
the average of the leaked information 
${\mathbb{E}}_F  \text{S}_{\text{strong}}(M;Z^n)$ goes to zero exponentially.

Since this method can be applied to any algebraic error correcting code,
our method can be applied to the case when the code $(\phi_{e,n},\phi_{d,n})$ is a LDPC code or a Polar code.
So, our method provides an efficient wire-tap code based on LDPC codes and Polar codes,
which attains the wire-tap capacity for symmetric channel
because the maximum in \eqref{eq:secrecy_capacity} is attained when $P_X$ is the uniform distribution.

Note that these inequalities still hold even for any input distribution for $M$ 
and any asymmetric channel $W_{Z|X}$.
When $W_{Z|X}$ is symmetric.
$\text{S}_{\text{strong}}(M;Z^n)$ with a general input distribution for $M$
is upper bounded by that with the uniform distribution for $M$.
So, we can guarantee the semantic secrecy.
In this case, $q_X$ is also chosen to be the uniform distribution for $X$.
However, when $W_{Z|X}$ is asymmetric, we need to 
expurgate the messages to guarantee the semantic secrecy.
This process requires large calculation complexity.

Further, even when the auxiliary random number $L_n$ is not uniform,
a similar evaluation is available by slightly different coding given in \cite[Section XI]{Hayashi2012}.
Then, the above security evaluation holds with the replacement of $2^{-s k_n'} $ 
by $e^{s H_{1+s}(L_n)}$,
where $H_{1+s}(X)$ is the R\'{e}nyi entropy defined as
\begin{align}
H_{1+s}(X):=-\frac{1}{s}\log\sum_{x}P_X(x)^{1+s}  .
\end{align}
These evaluations still hold even when the output system of $W_{Z|X}$ is continuous,
e.g., the AWGN channel \cite[Appendix D]{Hayashi2012}.
In particular, the AWGN channel with binary input can be regarded as a symmetric channel.

\section{Satellite wiretap channel model}
\label{sec:app-sat-chan}

Our general construction assumes $n$ satellite channel uses. We analyse the wiretap secrecy of a typical satellite channel with $2^m$-level phase shift keying ($2^m$-PSK) modulation with additive white Gaussian (AWGN) noise, which can be unfaded or faded. 

\subsection{Transmitted signal model}

The bandpass $i$-th $2^m$-PSK signal,
for $i=1,\ldots,2^m$, can be mathematically represented as 
\[
x_{i}^{BP}\left(t\right)=\sqrt{2P}Re\left\{ x_{i}^{LP}\left(t\right)e^{j2f_{c}\pi t}\right\} ,
\]
where $x_{i}^{BP}\left(t\right)$ is the continuous complex low pass
equivalent signal (or complex envelope) that conveys the complex information
symbol, 
\[
x_{i}^{LP}\left(t\right)=\left[X_{i}^{Re}+jX_{i}^{Im}\right]p\left(t\right)=X_{i}p(t),
\]
which for $2^m$-PSK is given as
\[
X_{i}=\exp\left[{j\frac{2\pi\left(i-1\right)}{2^m}}\right],\begin{array}{cc}
 & i=1,\ldots,2^m\end{array}.
\]

The signal $p(t)$ is the shaping pulse of duration equal to the symbol
duration, $T_{s}$. The factor $\sqrt{2P}$ represents the peak value of the sinusoidal carrier with power dissipated $P$ (over unitary load). Note that we choose the complex baseband signal representation to be normalized with $E_X=E\left[\left|X_{i}\right|^{2}\right]=1$. 

Geometrical representation in the signal space is useful for secrecy analysis.
For example, for $m = 1$ we have the binary signal called BPSK. Assuming $b(t)$ takes the value +1 when binary  $"1"$ is to be transmitted and the value -1 when binary  $"0"$ is to be transmitted, we can represent the BPSK signal as two points located on a single orthonormal carrier
\[
x^{BPSK}\left(t\right)=b\left(t\right)\sqrt{E_b}\phi\left(t\right),
\]
with one point located at $+\sqrt{E_b}$ and another point located at  $-\sqrt{E_b}$ with $E_b=PT_b$, $T_b$ is the bit period and the orthonormal carrier is  
\[
\phi\left(t\right)=\sqrt{2/T_b}cos\left(2f_{c}\pi t\right) .
\]

This signal space view considers the set of waveforms as the vector space of finite-energy complex functions, the (infinite-dimensional) Hilbert space, ${\cal L}_2$. Hence, this view is related to the (continuous) transmission domain (e.g time). Alternatively, the signal space view representing the complex low pass equivalent signal is convenient for analysis of geometrical properties of the (digital) information. In this case, the transmission domain can be abstracted away.

\subsection{Large-scale model}

It is convenient to characterise the received signal model in terms
of large-scale and short/medium-scale signal variations for both Bob
and Eve. 

For the large-scale, the low pass complex statistical signal models of the
received signals after demodulation during a symbol interval are

\begin{equation}
\begin{split}
Y^{l}
&=h_{Y}^{l}\left(\pi_{B},\lambda_{c}\right)X+N_{B}\\
Z^{l}
&=h_{Z}^{l}\left(\pi_{E},\lambda_{c}\right)X+N_{E},
\end{split}
\end{equation}

where $X$ is the complex transmitted symbol. $h_{Y}^{l}\left(\pi_{Y},\lambda_{c}\right)$ and $h_{Z}^{l}\left(\pi_{Y},\lambda_{c}\right)$
are deterministic coefficients that describe the (square root of)
the power decay due to propagation. $\pi_{B}=\left(\rho_{B},\theta_{B}\right)$
and $\pi_{E}=\left(\rho_{E},\theta_{E}\right)$ describe Bob's and
Eve's polar coordinates (with Alice's location as reference of coordinate system), where $\rho_{B}$
is the distance from Alice to Bob and $\theta_{B}=0$. $\lambda_{c}$
is the transmission wavelength, i.e. $\lambda_{c}=c/f_{c}$ with $c$
the speed of light. 

The dependency of the propagation attenuation
with the wavelength is only due to how the system parameters are
measured for computing the free loss propagation. The coefficients
$N_{B}$ and $N_{E}$ are complex circular Gaussian random variables
with zero mean and spectral density $n_{B}$ and $n_{E}$, respectively.

For the purpose of the security analysis to be made here, without
loss of generality, it is sufficient to consider a simplified e.g.  uplink budget between Alice and Bob that captures the wiretap large-scale
system geometry. In this case, the deterministic coefficient for Bob can be
expressed as 
\begin{align}
h_{Y}^{l}\left(\pi_{B},\lambda\right)=\sqrt{g_{A,max}g_{B,max}}\frac{1}{\rho_{B}}\left(\frac{\lambda_{c}}{4\pi}\right),
\end{align}

where $g_{A,max}$ and $g_{B,max}$ are the gains of Alice's antenna
and Bob's antennas to each other direction. We assume perfect polarization
antenna matching and we omit additional transmission/reception path
losses and (geo-climatic) atmospheric contributions as a first approximation.
For reliable communication, the satellite system design should ensure
a minimal received power level at the legitimate receiver so that
the link budget is closed. 

Now denote the distance between Alice and Eve as $\rho_{E}$ and the
angle in degrees from the maximum directivity direction from Alice's antenna
to Eve as $\theta_{E}$. Let's introduce the parameter $\alpha\left(\theta_{E}\right)$
to account for spatial attenuation due to Alice's antenna radiation's
pattern with respect to the Bob's boresight angle. $\alpha\left(\theta_{E}\right)$
is known because emission of radiation in space communication is regulated.
Let's also introduce the parameter $\mu\left(\theta_{E}\right)$ to
account for the relative antenna gain between Bob and Eve, i.e. $\sqrt{g_{E}\left(\theta_{E}\right)}=\mu\left(\theta_{E}\right)\sqrt{g_{B,max}}$. with $\mu$ taking values between 0 and 1.
Finally, we define $\beta\left(r,\rho_{E}\right)$ to account for relative
propagation losses between Bob and Eve as
\begin{align}
\beta^2\left(r,\rho_{E}\right)=\frac{\rho_{B}^{2}}{\rho_{E}^{r}}.
\end{align}
The exponent $r$ accounts for the power attenuation decay that affects Eve's propagation channel. Different values of the exponent model
correspond to different assumptions about Eve. For example, Eve can be modelled as a terrestrial, aerial or satellite station. For example, while for the satellite case $r=2$, in case of aerial Eve, a good
assumption is to consider a large scale two-ray ground multipath model,
with $r>2$. We can then write 
\begin{align}
h_{Z}^{l}\left(\pi_{E},\lambda\right)=h_{Y}^{l}\left(\pi_{B},\lambda\right)\alpha\left(\theta_{E}\right)\mu\left(\theta_{E}\right)\beta\left(r,\rho_{E}\right).\label{hlz}
\end{align}

\subsection{Medium/short-scale model}

For the medium/short-scale, the received
signals after demodulation for a symbol interval are
\begin{equation}
\begin{split}
Y
&=h_{Y}\left(\pi_{B},\lambda_{c}\right)X+N_{B}\\
Z
&=h_{Z}\left(\pi_{E},\lambda_{c}\right)X+N_{E},
\end{split}
\end{equation}
where $h_{B}\left(\pi_{B},\lambda_{c}\right)$ and $h_{E}\left(\pi_{E},\lambda_{c}\right)$
are now random variables that describe the medium/short term variations
normalised with respect to the deterministic coefficients $h_{B}^{l}\left(\pi_{B},\lambda_{c}\right)$
and $h_{E}^{l}\left(\pi_{E},\lambda_{c}\right)$, respectively. The
first-order statistical characterisation of these random variables for Bob
highly depends on the frequency and the satellite mission. For Eve, it depends on the type of the relative location and whether
or not transmission is affected by atmospheric losses (e.g. Eve is below/above
the clouds). Both for Bob and Eve, for the cases when there is fading due to multipath propagation (short-scale channel dynamics), it is usually modelled separated from shadow fluctuations (medium-scale channel dynamics). The second-order statistical characterisation mostly depends on the relative speeds. 

\subsection{Comparison with spread spectrum signal model}
It is clarifying to identify here the difference between the signal model for physical layer security based on spread spectrum and on wiretap coding. In order to see this, note that the data signal stochastic process generated by Alice is  
\[
a^{LP}\left(t\right)=\sum_{j}a_{j}^{LP}\left(t\right)=\sum_{j}X_{j}p\left(t-jT_{s}\right),
\]
where $X_{j}$ are random variables taking $2^m$-PSK values. When
spread spectrum techniques are used, the data signal is further modulated
using a pseudo-random or cryptographic pseudo-random sequence $c_{PN}(t)$
either using frequency hopping or direct multiplication. In the latter
case the data signal is 
\begin{align}
a^{SS}\left(t\right)=a^{LP}\left(t\right)c_{PN}(t),
\end{align}
with 
\begin{align}
c_{PN}(t)=\sum_{k}C_{k}g\left(\frac{t-kT_{c}}{T_{c}}\right),
\end{align}
where $g(t)$ is the chip waveform assuming support limited to one
chip interval. The chip duration, $T_{c}$, is assumed as $T_{c}=T_{s}/G$,
with $G$ the number of chips per symbol interval. 

From the signal model we can see that physical layer security based
on spread spectrum operates at waveform level. The ``plaintext''
is the information signal (here represented in complex lowpass equivalent
form), which is encrypted into a noise-like waveform using a (possibly
crypto) pseudo-random sequence. Differently, the information-theoretic
physical layer security discussed here operates at information signal
level. Now, the ``plaintext'' is the $k_{n}$-length information
message that Alice wants to transmit to Bob. Both security mechanisms could in principle operate
jointly. However, we do not discuss such possible joint design here.

\subsection{Satellite Gaussian channel model} 

A simple but relevant satellite channel is the Gaussian channel, i.e. we assume the channel has no fading
\begin{align}
h_{Y}\left(\pi_{B},\lambda_{c}\right)=1.
\end{align}

In the following, we assume some fixed frequency so we can drop the dependency of the wavelength. Assuming the deterministic channel coefficient known to transmitter and legitimate and illegitimate receivers, the short/medium term signal model of the Gaussian wiretap satellite channel is
\begin{equation}
\begin{split}\label{eq:Gaussig1}
Y&=X+N_{B}\\
Z&=\gamma_gX+N_{E},
\end{split}
\end{equation}
where the deterministic coefficient $\gamma_g $ is defined as
\begin{align}
\gamma_g : = \alpha\left(\theta_{E}\right)\mu\left(\theta_{E}\right)\beta\left(r,\rho_{E}\right).\label{grrt}
\end{align}

We can re-write the model so that Bob's signal is the reference for Eve's signal
\begin{equation}\label{eq:Gaussig2}
\begin{split}
Y&=X+\sqrt{n_{B}}X_1\\
Z&=\gamma_gX+\sqrt{\gamma_nn_{B}}X_1,
\end{split}
\end{equation}
where the coefficient $\gamma_n$ taking values between 0 and 1. This coefficient accounts for the relative performance of Eve's w.r.t. Bob's receivers in terms of Gaussian noise power.  $X_1$ is a zero-mean circular complex Gaussian random variable with unit variance.

From the resulting Gaussian wiretap signal model we can see that the degradation of Eve's channel with respect to Bob's channel is driven by two components of the wiretap system:
\begin{itemize}
\item  \textbf{Geometrical properties}. The geometry of the large-scale wiretap satellite scenario introduces degradation due to the relative attenuation in Eve's channel with respect to Bob's channel. This degradation is captured by $\beta\left(r,\rho_{E}\right)$,
\item  \textbf{System parameters}. These include Eve's receiver antenna gain relative to Bob's, which is captured by $\mu\left(\theta_{E}\right)$, Alice's transmitter antenna gain relative to Eve's, captured by $\theta\left(\theta_{E}\right)$ and the relative receivers performance, captured by $\gamma_n$. 
\end{itemize}

In particular we see that Eve's signal amplitude is degraded w.r.t. Bob's if $\gamma_g<1$. We also see that Gaussian variability is bigger for Eve's signal whenever  $\gamma_n>1$. Consequently,  the geometrical influence in the signal amplitude could compensate for the effect of Eve's system parameters. In particular, Eve could have stronger signal than Bob if 
\begin{equation}\label{eq:strongeve}
 \alpha\left(\theta_{E}\right)\mu\left(\theta_{E}\right)>1/\beta\left(r,\rho_{E}\right).
\end{equation}

In order to gain further insight, assume that the power decay represented by the parameter $\alpha(\theta_E)$ is exponential as  $\alpha(\theta_E) = 1/\theta^a$.   According to \eqref{eq:Gaussig2} and assuming equally good antennas for Bob and Eve, the angular spatial condition for having Eve's signal amplitude below Bob's as a function of distance proximity is
\[
\theta_{E}>\left(\rho_{B}/\rho_{E}^{r/2}\right)^{a/2}.
\]

\begin{figure}[tbh]
\includegraphics[scale=0.23]{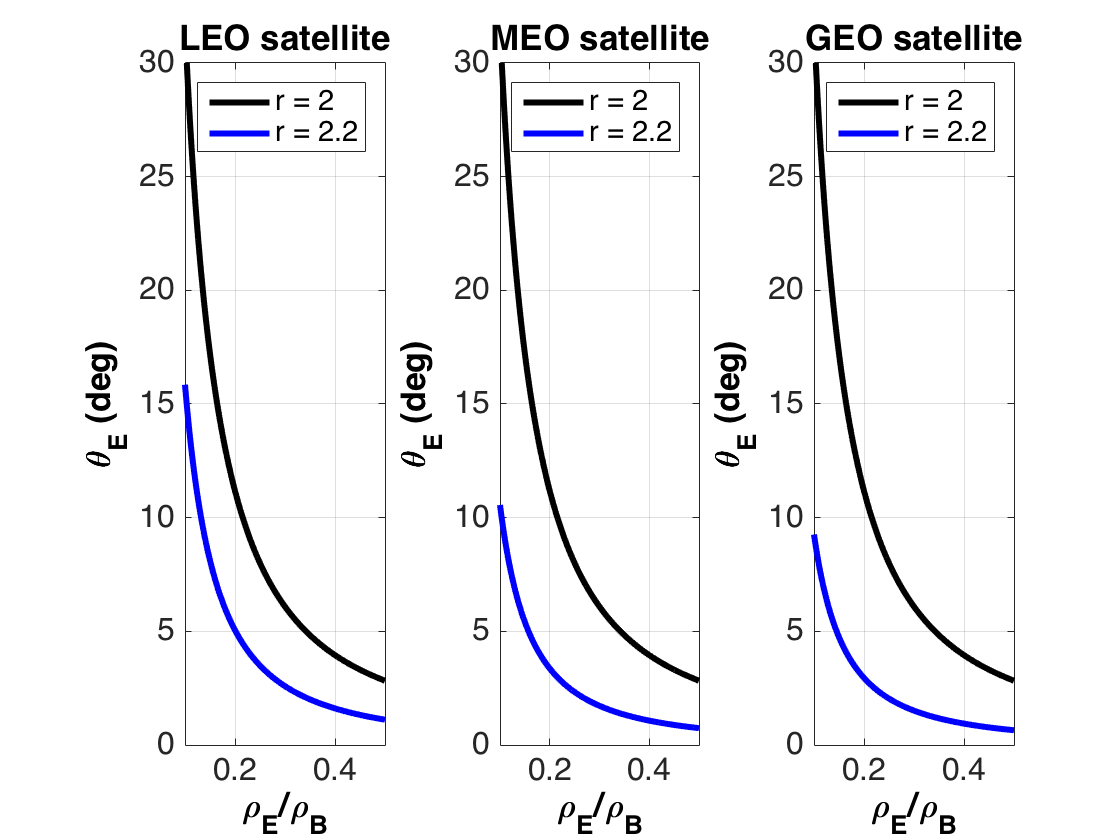}
\protect\caption{Illustration of Eve's closest angular distance to Alice (as a function of relative distances) so Eve's signal amplitude is $\gamma_g<1$. }
\end{figure}

It is observed that for equal (free space) propagation conditions for Bob and Eve, the angular proximity does not depend on the absolute distances. This means that the same condition holds for any type of satellite orbit. This fact is illustrated in Figure 1, where for $r=2$ the same minimum angular proximity holds for the three considered types of satellite. It is also shown how angular proximity depends on the type of orbit whenever Eve's propagation conditions are worse than Bob's. In such case, Eve must be closer to the satellite link. 

\begin{figure}[tbh]
\includegraphics[scale=0.23]{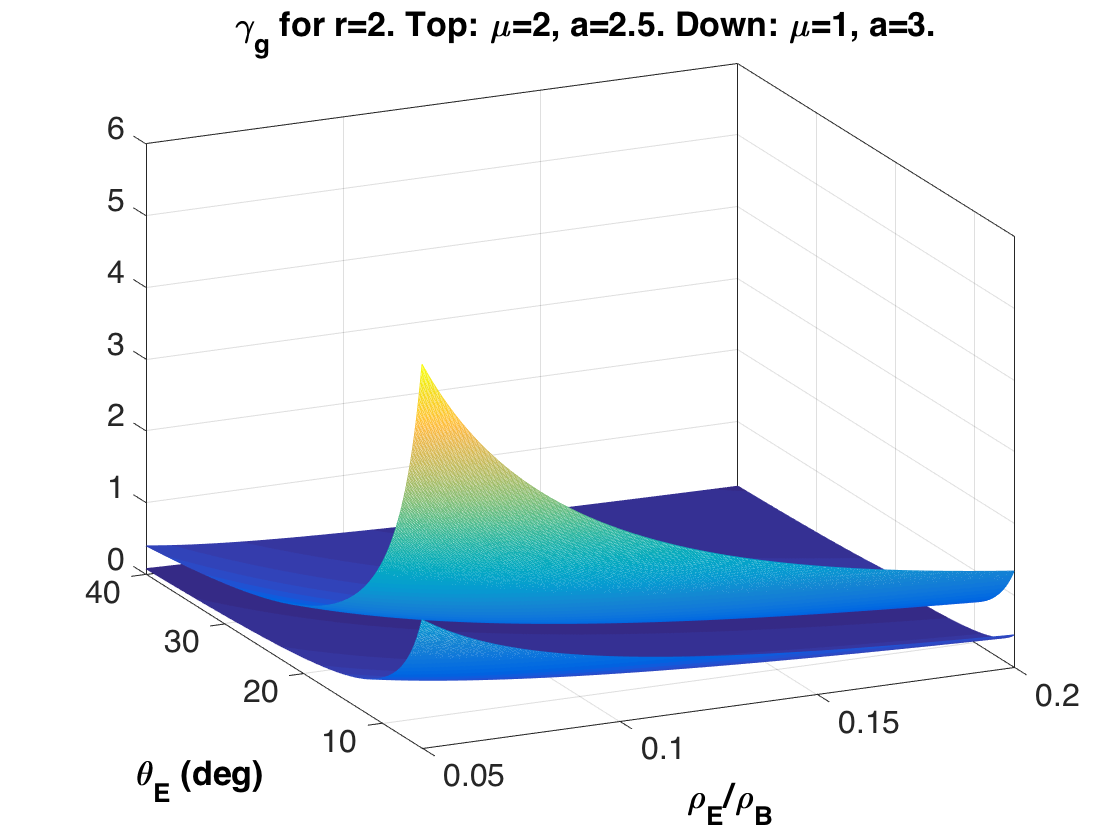}
\protect\caption{Illustration of the influence of wiretap system parameters on Eve's amplitude signal degradation for different values of system parameters.}
\end{figure}

\begin{figure}[tbh]
\includegraphics[scale=0.23]{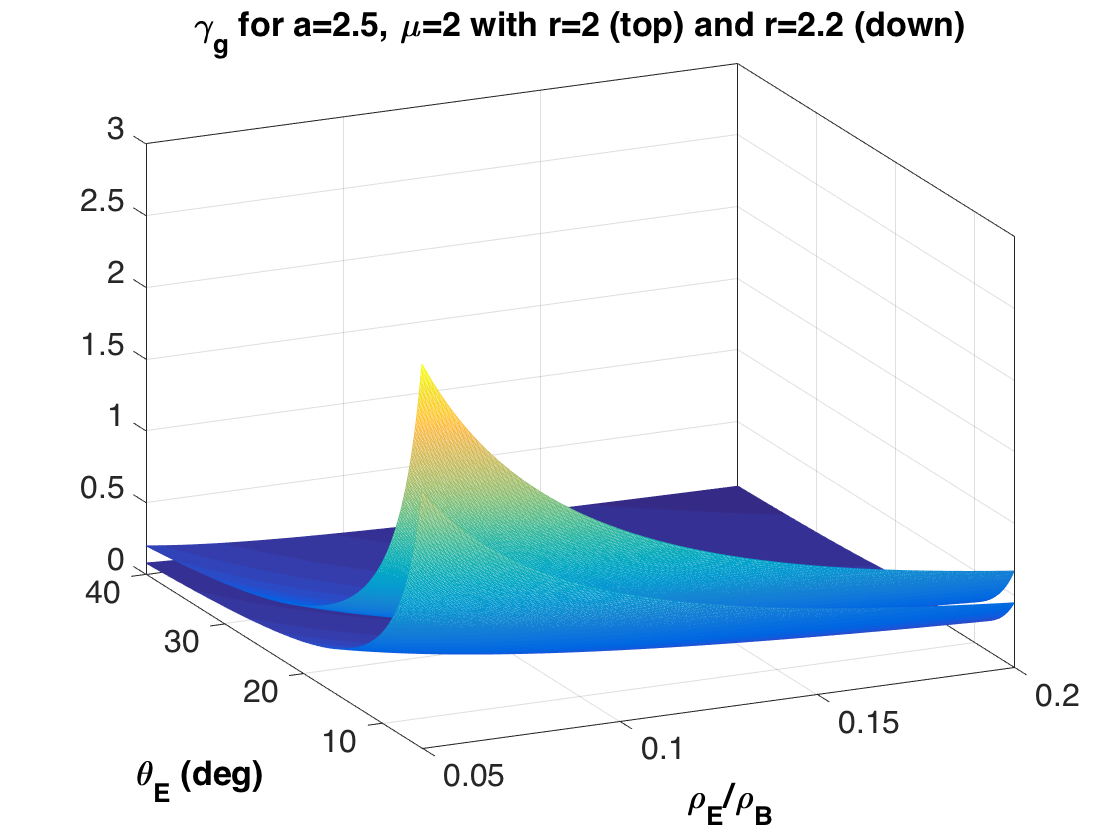}
\protect\caption{Illustration of the influence of wiretap channel propagation on Eve's amplitude signal degradation for different propagation laws. }
\end{figure}

Figure 2 illustrates two numerical examples on the amount of energy that can be captured by Eve as given by $\gamma_g$. It shows the case where both Bob and Eve have equal (free space) propagation law. It can be seen that when Eve is closer to Alice than Bob both in distance and in angle (represented by small $\rho_{E}/\rho_{B}$ and small $\theta_{E}$, respectively) the quality of Eve's channel is always better (i.e. $\gamma_g>1$). However, usually the proximity of satellite stations more prone to wire-tapping (e.g. a space data link) is in general protected as an asset and Eve will not be able to get too close. The upper surface in the figure shows a situation beneficial for Eve due to advantageous transmitter and receiving antenna patterns, showing the significant influence of diagram antenna patters.The lower surface shows the corresponding disadvantageous situation for Eve only to due to disadvantageous transmitter and receiving antenna patterns. It can be seen that in th
 is case 
 Eve needs to be extremely close to Alice for  $\gamma_g>1$.

Figure 3 illustrates the effect of a propagation law for Eve of $r=2.2$ assuming LEO orbit. As expected, it can be seen it is less beneficial for Eve than the free-space propagation law. However, we also saw that Eve could compensate this effect with an optimized choice of system parameters.
 
The above conditions are simplified as we have not taken into account atmospheric losses and accurate propagation law for Eve. This will depend on the aerial vehicle used by Eve. However, it is clear that given concrete system and propagation assumptions for Bob and Eve, it is possible to compute with high accuracy the large scale geometrical region where  $\gamma_g<1$ holds. 

\section{Secrecy analysis: infinite-length regime}
\begin{figure}[tbh]
\includegraphics[scale=0.55]{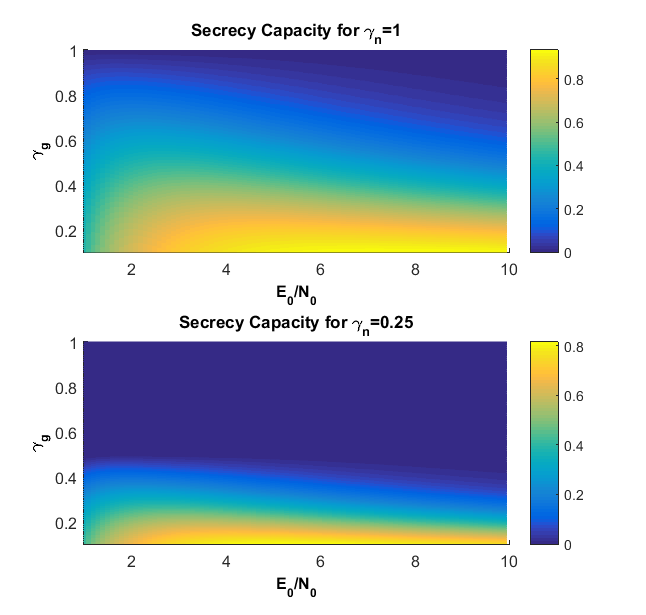}
\protect\caption{Secrecy capacity showing the effect of Eve's noise.}
\label{F14}
\end{figure}

\begin{figure}[tbh]
\includegraphics[scale=0.5]{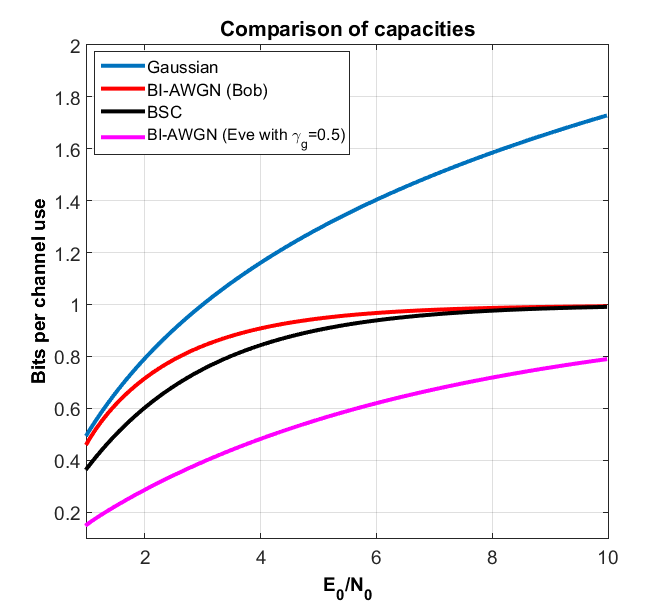}
\protect\caption{Comparison of BI-AWGN Bob's and Eve's capacities with known Gaussian and BSC capacities. Eve's capacity is shown for the case $\gamma_g=0.5$}
\end{figure}

\begin{figure}[tbh]
\includegraphics[scale=0.5]{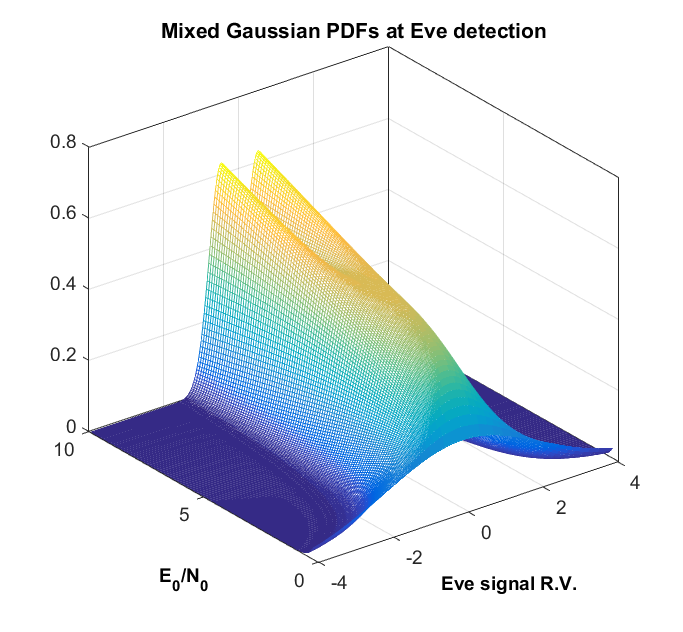}
\protect\caption{Mixed Gaussian probability density functions at Bob's detection.}
\end{figure}

\begin{figure}[tbh]
\includegraphics[scale=0.5]{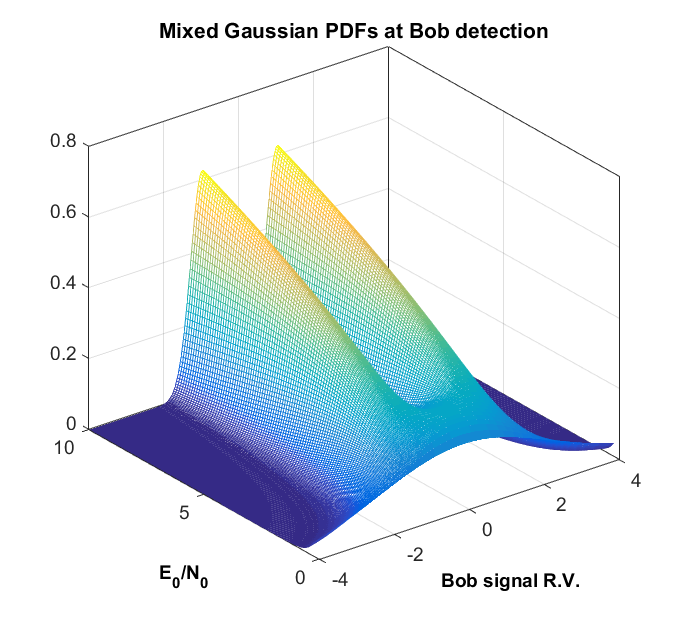}
\protect\caption{Mixed Gaussian probability density functions at Eve's detection for the case $\gamma_g=0.5$.}
\end{figure}

\begin{figure}[tbh]
\includegraphics[scale=0.55]{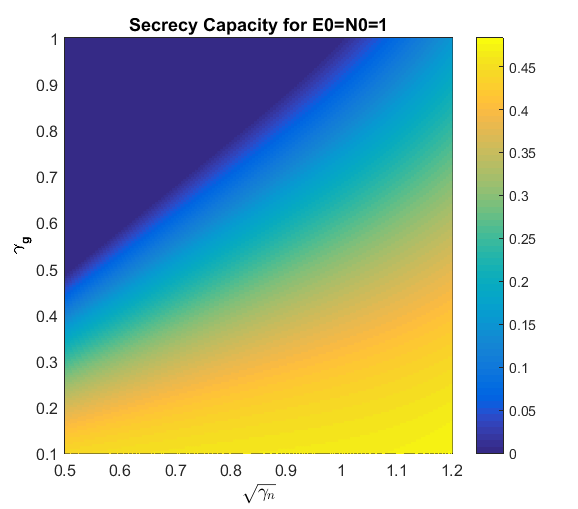}
\protect\caption{Visualization of the condition for positive capacity.}
\end{figure}
In this regime we use standard information theoretical terminology
and arguments.
The secrecy metrics in this regime is information-theoretic strong
secrecy and the implied asymptotic semantic security.

To do the analysis, we first map our wiretap Gaussian satellite channel
model to the information-theoretic wiretap channel model. 
The way to do it is to identify the satellite channel with a two-user memoryless broadcast channel, with the legitimate transmitter having input $X$
and two receiver observations $Y$ and $Z$, the legitimate and intruder
respectively.

From the signal models, we can now identify the wiretap communication
channel, i.e. the statistical relation between the transmitted signal
and the received signal during $n$ available channel uses. This channel
is statistically described by the conditional distribution $W_{YZ|X}$.
We saw that the secrecy capacity of a discrete memoryless wiretap
channel is completely characterized for the weak secrecy, strong secrecy,
and semantic secrecy metrics, as \eqref{eq:secrecy_capacity}. We
also remember that Bob's and Eve's observed signal are given in \eqref{eq:Gaussig2}.

The independent Eve's and Bob's channels are discrete input and continuous
AWGN output channels. The input random variable $X$ takes vales in
${\cal X}:=\left\{ \exp\left[{j\frac{2\pi\left(i-1\right)}{2^{m}}}\right],i=1,\ldots,2^{m}\right\} $,
which is called $2^{m}$-PSK modulation.

According to our satellite channel model, we can express the secrecy
capacity as 
\begin{align}
C_{s}=\underset{q_{X}}{\max}\left(I(Y;X)-I(Z;X)\right)_{+},\label{eq:Cs_max}
\end{align}
which depends on the constant coefficients $\gamma_{g}$, and $\gamma_{n}$
introduced in Section 5, which represent the degradation of Eve's
signal amplitude and noise power w.r.t. Bob's, respectively.

The mutual information for Bob's channel is 
\begin{align*}
I(Y;X)=\int_{-\infty}^{\infty}\sum_{x\in{\cal X}}W_{Y|X}\left(y|x\right)q_{X}\left(x\right)\log_{2}\frac{W_{Y|X}\left(y|x\right)}{W_{Y}\left(y\right)}dy
\end{align*}
where 
\begin{align*}
W_{Y}\left(y\right):=\sum_{x}W_{Y|X}\left(y|x\right)q_{X}\left(x\right).
\end{align*}

The mutual information for Eve's channel is 
\begin{align*}
I(Z;X)=\int_{-\infty}^{\infty}\sum_{x\in{\cal X}}W_{Z|X}(z|x)q_{X}(x)\log_{2}\frac{W_{Z|X}(z|x)}{W_{Z}(z)}dz
\end{align*}
where 
\begin{align*}
W_{Z}(z):=\sum_{x}W_{Z|X}(z|x)q_{X}(x).
\end{align*}

In this case, the maximization in \eqref{eq:Cs_max} is achieved by
the uniform distribution due to the symmetry. 
In general, the bit error probability of $2^{m}$-PSK is difficult
to obtain for an arbitrary integer $m$. We assume from now $m=1$
so that $X$ takes values in $\{-1,1\}$, which is called BPSK modulation.
The secrecy capacity for BPSK input corresponds to the positive values
of 
\begin{align*}
  C_{s}=&\frac{1}{2}\int W_{Y|X}\left(y|1\right)\log_{2}\frac{W_{Y|X}\left(y|1\right)}{W_{Y}\left(y\right)}dy\\
 & +\frac{1}{2}\int W_{Y|X}\left(y|-1\right)\log_{2}\frac{W_{Y|X}\left(y|-1\right)}{W_{Y}\left(y\right)}dy\\
 & -\frac{1}{2}\int W_{Z|X}\left(z|1\right)\log_{2}\frac{W_{Z|X}\left(z|1\right)}{W_{Z}\left(z\right)}dz\\
 & -\frac{1}{2}\int W_{Z|X}\left(z|-1\right)\log_{2}\frac{W_{Z|X}\left(z|-1\right)}{W_{Z}\left(z\right)}dz.
\end{align*}

We assume that\footnote{Usually for a bandpass $2^{m}$-PSK the assumption is $\sigma_{B}^{2}=N_{0}/2$.
However, since we focus on a BPSK signal it is coherent to assume
instead $\sigma_{B}^{2}=N_{0}$ as the two BPSK symbols are real.} $n_{B}=\sigma_{B}^{2}=N_{0}$. The conditional probabilities for
Bob's observation are 
\begin{align*}
W_{Y|X}\left(y|1\right)=\frac{1}{\sqrt{2\pi N_{0}}}e^{-\frac{\left(y-1\right)^{2}}{2N_{0}}},\\
W_{Y|X}\left(y|-1\right)=\frac{1}{\sqrt{2\pi N_{0}}}e^{-\frac{\left(y+1\right)^{2}}{2N_{0}}},
\end{align*}
while the conditional probabilities for Eve's observation are 
\begin{align*}
W_{Z|X}\left(z|1\right)=\frac{1}{\sqrt{2\pi\gamma_{n}N_{0}}}e^{-\frac{\left(z-\gamma_{g}\right)^{2}}{2\gamma_{n}N_{0}}},\\
W_{Z|X}\left(z|-1\right)=\frac{1}{\sqrt{2\pi\gamma_{n}N_{0}}}e^{-\frac{\left(z+\gamma_{g}\right)^{2}}{2\gamma_{n}N_{0}}}.
\end{align*}

This secrecy capacity is asymptotically achievable by the wiretap
code given in Section \ref{sec:chann-resolv-with}. 
The calculation of this capacity can be computed numerically 
in Figure \ref{F14}.

To see the detail structure of the secrecy capacity, 
we calculate Bob's and Eve's capacities 
\begin{align*}
  C^{B}=&\frac{1}{2}\int W_{Y|X}\left(y|1\right)\log_{2}\frac{W_{Y|X}\left(y|1\right)}{W_{Y}\left(y\right)}dy\\
 & +\frac{1}{2}\int W_{Y|X}\left(y|-1\right)\log_{2}\frac{W_{Y|X}\left(y|-1\right)}{W_{Y}\left(y\right)}dy \\
  C^{E}=&
 \frac{1}{2}\int W_{Z|X}\left(z|1\right)\log_{2}\frac{W_{Z|X}\left(z|1\right)}{W_{Z}\left(z\right)}dz\\
 & +\frac{1}{2}\int W_{Z|X}\left(z|-1\right)\log_{2}\frac{W_{Z|X}\left(z|-1\right)}{W_{Z}\left(z\right)}dz
\end{align*}
as Figure 5 when $\gamma_g=0.5$. 
Here, 
we need to treat the mixed Gaussian probability density functions $W_{Y}\left(y\right)$ and $W_{Z}\left(z\right)$.
These probability density functions are illustrated in Figures 6 and 7.

This capacity is known as the capacity of the Binary-Input AWGN (BI-AWGN) channel and is illustrated in Figure 5. It is shown in comparison with the well known capacities of a Binary Symmetric Channel (BSC) and the Gaussian channel. Figure 6 illustrates the effect of channel degradation on the mixed Gaussian pdfs for the case of $\gamma_g=0.5$. It is observed that Eve's distinguishability of Gaussians drops w.r.t to Bob's due to channel degradation.

The positivity of the secrecy capacity can be analyzed as follows.
When the new random variable $Z':= \frac{Z}{\sqrt{\gamma_n}}$,
the conditional probabilities for Eve's observation are 
\begin{align*}
W_{Z'|X}\left(z'|1\right)&=\frac{1}{\sqrt{2\pi N_{0}}}e^{-\frac{\left(z-\frac{\gamma_{g}}{\sqrt{\gamma_{n}}}\right)^{2}}{2N_{0}}},\\
W_{Z'|X}\left(z'|-1\right)&=\frac{1}{\sqrt{2\pi N_{0}}}e^{-\frac{\left(z+\frac{\gamma_{g}}{\sqrt{\gamma_{n}}}\right)^{2}}{2N_{0}}}.
\end{align*}
Comparing $W_{Y|X}$ and $W_{Z'|X}$,
we find that 
$C^B>C^E$ if and only if
\begin{align}
\gamma_g < \sqrt{\gamma_n}.\label{12-18}
\end{align}
because the channel $W_{Z'|X}$ is given as a degraded channel 
of $W_{Y|X}$ only in the case \eqref{12-18}.

Now, we see this necessarily and sufficient condition from the viewpoint of \textit{c-separation} of a mixed Gaussian density function introduced in \cite{Dasgupta99}. Two Gaussians $N(\nu_1,\sigma^2)$ and  $N(\nu_2,\sigma^2)$ are c-separated if $||\nu_1-\nu_2||\geq c\sigma$. In particular a 2-separated mixture corresponds roughly to almost completely separated Gaussians. In our case, assuming a 2-separation condition for Eve's mixed Gaussian $p_{mix}\left(z|\gamma_g,\gamma_n\right)$, would correspond to Eve being able to distinguish between the two symbols. Hence, an approximate condition of positive secrecy capacity is to assume Eve's mixed Gaussian to be less than 2-separated, which yields the following condition
\[
\frac{\gamma_g}{\sqrt{\gamma_n}} < \frac{\sqrt{N_0}}{E_0}.
\]

The accuracy of this condition is illustrated in Figure 8. It shows the case of $E_0=N_0=1$, in which case the condition for positive capacity is \eqref{12-18}.

\section{Security analysis: finite-length regime}

The wiretap code must provide reliable communication with secrecy guarantees to a specific system or service operating in the finite-length regime.
In the following, we restrict our error correction and the hash functions to linear operations.
In this case $\max_{p_M}\text{S}_{\text{strong}}(M;Z^n)$ is realized when $p_M$ is the uniform distribution
because our wiretap satellite $W_{Z|X}\left(z|x\right)$ is symmetric.
So, we simply denote this leaked information with the uniform distribution by
$\text{S}_{\text{strong}}(M;Z^n)$ 
Since $\max_{p_M}\text{S}_{\text{strong}}(M;Z^n)$
upper bounds the semantic secrecy with a proper coefficient,
we adopt $\text{S}_{\text{strong}}(M;Z^n)$
as our secrecy metric in this regime.
Since our wiretap satellite $W_{Z|X}\left(z|x\right)$ is symmetric, 
its upper bound given in \eqref{H12-13}
is given as the case when $q_X$ is the uniform distribution. 
So, due to the symmetric property, the asymptotically attainability of the secrecy capacity also entails semantic secrecy attainability. 

As an evaluation of the leaked information,
we find that the average of the leaked information for our wiretap satellite channel is bounded as
\begin{align}
{\mathbb{E}}_F \text{S}_{\text{strong}}(M;Z^n)
\le &
 \text{S}_{\text{strong}}(s|M;Z^n)\nonumber \\
:=& \frac{1}{s} 2^{-s k_n'}e^{n E_{0,\max}(s|W_{Z|X})} \label{10-23}
\end{align}
for $s \in (0,1]$, where
$F$ is the hash function whose distribution is given in Section 4, and
\begin{align*}
E_{0,\max}(s|W_{Z|X})= 
\log \int (\sum_{x \in \{1,-1\}} 
\frac{1}{2} W_{Z|X}(z|x)^{\frac{1}{1-s}} )^{1-s} dz .
\end{align*}
In order to make explicit the dependency on the geometry and system parameters we express the exponent as
\begin{align*}
&E_{0,\max}(s|W_{Z|X}) \\
=& 
\log \int 
\Bigg(
\frac{1}{2}
\Bigg(
\left[\frac{1}{\sqrt{2\pi N_{0}}}e^{-\frac{\left(z-\gamma_g\right)^{2}}{2\gamma_nN_{0}}}\right]^\frac{1}{1-s} \\
&+
\left[\frac{1}{\sqrt{2\pi N_{0}}}e^{-\frac{\left(z+\gamma_g\right)^{2}}{2\gamma_nN_{0}}}\right]^\frac{1}{1-s}
\Bigg)
\Bigg)^{1-s} dz .
\end{align*}

The average of the leaked information ${\mathbb{E}}_{F}  \text{S}_{\text{strong}}$ goes to zero exponentially when $\frac{E_{0,\max}(s|W_{Z|X})}{s}$ is smaller than $(\frac{k_n'}{n})\log 2$
with a real number $s \in (0,1]$.
This implies that 
when we randomly choose $F$, 
the quantity $\text{S}_{\text{strong}}(M;Z^n)$ is sufficiently small with high probability.

\begin{figure}[tbh]
\includegraphics[scale=0.45]{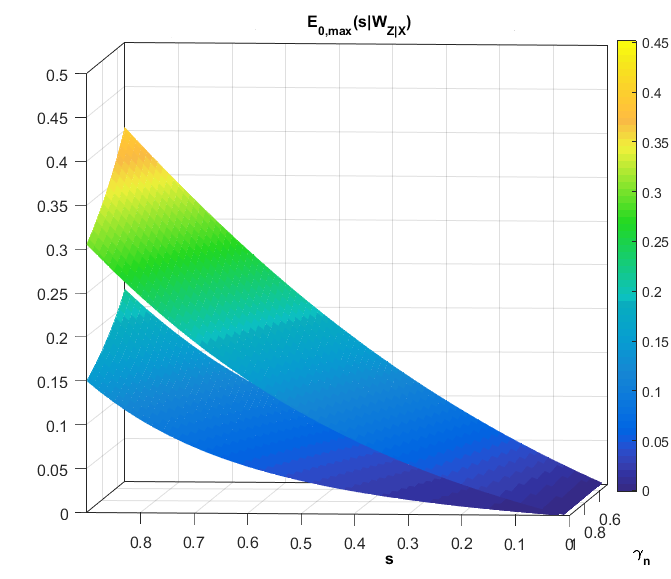}
\protect{\caption{Effect of the noise at Eve's receiver in the function $E_{0,\max}(s|W_{Z|X})$  for $\gamma_g = 0.6$ (top) and $\gamma_g = 0.3$ (down),
where $E_0=N_0=1$.}}
\end{figure}

For the finite-length numerical analysis, we make use of the range of values obtained in the infinite-length analysis for which the capacity is positive. For example, we can analyse operational points such that $\left (\gamma_g,\gamma_n>\gamma_g^2\right)$. For easier visualization we choose $E_0=N_0=1$. 

Figure 9 shows the function $E_{0,\max}(s|W_{Z|X})$ illustrating the effect of the noise at Eve's receiver for two operational points of positive capacity with  $E_0=N_0=1$. We can see that larger $\gamma_g$ leads to a bigger value of the exponent. Hence, Eve's leakage will consume more air time for security out of the overall budget of airtime for information and redundancy for reliability.

\begin{figure}[tbh]
\includegraphics[scale=0.45]{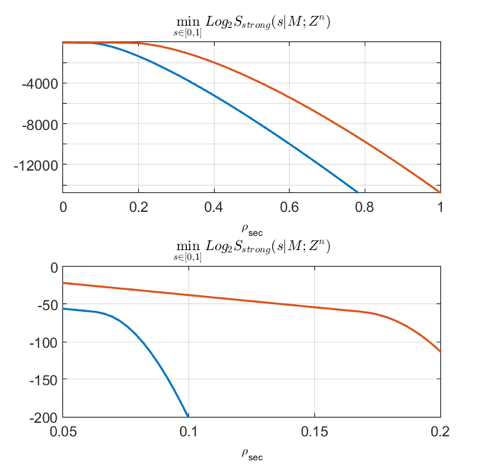}
\protect{\caption{
Secrecy performance at operative points $\left (\gamma_g,\sqrt{\gamma_n}\right)=\left(0.3,\sqrt{2}\right)$ (blue) and
$\left (\gamma_g,\sqrt{\gamma_n}\right)=\left(0.5,\sqrt{2}\right)$ (red) for the DVB-S2X system. The bottom figure is a zoom in for small wiretap coding rate.
}}
\end{figure}

\begin{figure}[tbh]
\includegraphics[scale=0.47]{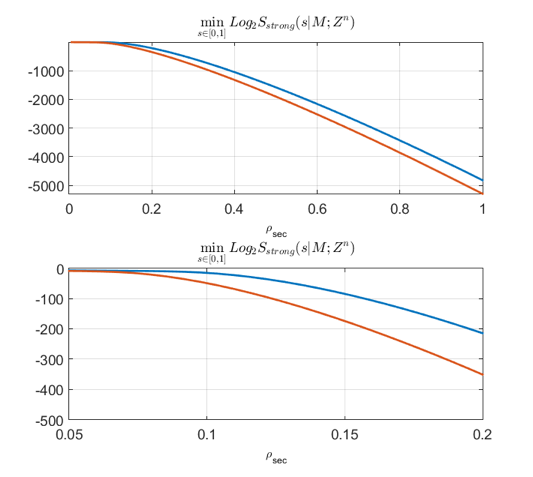}
\protect{\caption{
Secrecy performance at operative points $\left (\gamma_g,\sqrt{\gamma_n}\right)=\left(0.3,1\right)$ (red) and
$\left (\gamma_g,\sqrt{\gamma_n}\right)=\left(0.3,\sqrt{2}\right)$ (blue) for the Euclid system.The bottom figure is a zoom in for small wiretap coding rate.}}
\end{figure}

Now we analyse the performance that our wiretap code can provide in order to guarantee design target levels of simultaneous secrecy and reliability to a space channel.

We analyse two different space systems. First we assume a commercial satellite system using DVB-S2X air interface with quasi-error free goal. The theoretical satellite broadcast channel model BI-AWGN can be mapped to a typical DVB-S2X channel when using BPSK modulation and a LDPC code with medium coding block length of $n=32400$ and coding rates of 1/5, 11/45, 1/3. Adaptive coding and modulation would require secrecy rate to adapt accordingly.  A typical scenario in this case would be a broadcasting service over GEO satellite. Second, we assume the European Space Agency scientific mission called Euclid, planned to be launched in 2020 to map the geometry of the dark Universe. The scenario is very different since large quantity of scientific data will be transmitted through bandwidth-limited links with very long propagation delay. The CCSDS code LDPC AR4JA (8192, 4096) is foreseen for the K-band link (bit rate 73.84638 Mb/s), with a FER lower than $10^{-5}$.
Note that mission-specific services may prioritize reliability over secrecy or vice versa for the specific air interface. The secrecy analysis can be performed just considering the LDPC coding rate for each of the systems and analysing the trade off between reliability and security.

Let's denote the wiretap coding rate as $\rho_{sec}=k_n'/n$ and consider the LDPC coding rates for the above systems. Then, the coding rate budget available for information is limited by the wiretap coding rate that is set as design target.
Figure 10 illustrates the performance for DVB-S2X for different wiretap coding rates when the wiretap system operates at $\left (\gamma_g,\sqrt{\gamma_n}\right)=\left(0.3,\sqrt{2}\right)$ and $\left (\gamma_g,\sqrt{\gamma_n}\right)=\left(0.5,\sqrt{2}\right)$. 
The graphs show the minimum upper bound  $\min_{s \in [0,1]}\log_2 \text{S}_{\text{strong}}(s|M;Z^n)$ as a function of $\rho_{sec}$. The two examples are valid operation points that guarantee positive secrecy rate since in both cases we have the condition \eqref{12-18}. 
Now, if by design we assign $10\%$ of overall bit transmission budget for wiretap coding rate, the obtained secure coding rate has a guaranteed secrecy quantified as $\min_{s \in [0,1]}\log_2 \text{S}_{\text{strong}}(s|M;Z^n)=-200$ for the case of Eve being further away from Eve and $\min_{s \in [0,1]}\log_2 \text{S}_{\text{strong}}(s|M;Z^n)=-50$ when Eve is closer. We see that secrecy is consumed by the effect of the geometry represented by $\gamma_g$. From another perspective, provision of secrecy is an interplay between geometry and energy in the same way as provision of reliability. 
Figure 11 illustrates the performance for the Euclid system for different wiretap coding rates. Due to the characteristics of the system, now we observe the effect of different receiver performance at the eavesdropper. In this case we see that secrecy is consumed when Eve's has a better receiver than the legitimate receiver with respect to the case of equally performing receivers. 

In both cases, it is  clear that a design target of higher level of secrecy requires a higher amount of reliability redundancy to be sacrificed for secrecy. Hence, for any capacity achieving channel code that is in use today, it is possible to use a wiretap code on top of the channel code.  The tradeoff between guaranteed secrecy and overhead is equivalent to the tradeoff between guaranteed reliability and overhead. The difference now is that the secrecy overhead reduces the communication rate, which becomes the secrecy rate upper bounded by the secrecy capacity. However, the obtained reliable transmission can be guaranteed to be unconditionally secure for the geometry and system conditions that have been shown in Section 5. 
 
Note that for certain mission-specific air interfaces, the finite-length regime needs to be analyzed considering both the link and the physical layer framing. This is due to the fact that some air interface may define very short link layer frames and/or significant physical layer overhead. In such cases, the air interface overhead could be comparable to the coding overhead.

\section{Conclusions}

In this paper we have derived a realistic satellite wiretap channel model that we have mapped to information-theoretic wiretap models. The models allow to show that wiretap coding alone can protect from passive eavesdropping the channel-degraded areas of a (broadcasting in nature) satellite channel. The size of such protected areas can be evaluated and depends on the large-scale geometry of the satellite link and on system-dependent parameters such as the statistical channel models and the receiver noise and antenna patterns (which shape the broadcasting property of the channel). 

The performance of wiretap codes for the strong and semantic secrecy levels can be numerically quantified. Hence, for the wiretap protected areas it is possible to evaluate the secrecy capacity for the infinite-length regime or upper bounds for the finite-length regime. We have introduced a general wiretap code construction and a randomized hash function construction and illustrate the performance. In particular we illustrate the amount of bits of information that such wiretap code can transmit with unconditional security for realistic systems. Our results confirm the potential of wiretap coding as a physical layer security based on information theoretic metrics and principles.

From a system-level perspective, we have shown that our proposed wiretap satellite modeling and code design 
as well as secrecy analysis methodology are system-independent.
Hence, information-theoretic physical layer security allows a layered integration of wiretap coding as a sublayer of the physical layer. Such integration enables to incorporate additional information-theoretic security protocols running independently of the communication transmission.

\acknowledgments
The first author would like to acknowledge the European Space Agency for partially funding her work on physical layer security for satellite communications under contract SATNEX IV CoO1 WI2 No 4000113177/15/NL/CLP. In particular, she would like to thank Ignacio Aguilar Sanchez for helpful technical discussions.
The second author was supported in part by a MEXT Grant-in-Aid for Scientific Research (B) No. 16KT0017,
the Okawa Research Grant and Kayamori Foundation of Informational Science Advancement.

\bibliographystyle{IEEE}


\end{document}